\documentstyle[12pt]{article}

\def\ref#1{$^{#1)}$}
\def\Tr{{\rm Tr}}
\newcommand{\be}{\begin{equation}}
\newcommand{\ee}{\end{equation}}
\newcommand{\bea}{\begin{eqnarray}}
\newcommand{\eea}{\end{eqnarray}}
\newcommand{\Lag}{{\cal L}}
\newcommand{\superint}{\int \diff^{4}\theta}
\newcommand{\lowest}{|_{\theta =\bar{\theta}=0}}
\newcommand{\diff}{\mbox{d}}
\newcommand{\Diff}{{\cal D}}
\newcommand{\WaWa}{\Tr({\cal W}^{\alpha}{\cal W}_{\alpha})}
\newcommand{\WbWb}{\Tr({\cal W}_{\dot{\alpha}}{\cal W}^{\dot{\alpha}})}
\newcommand{\DaDa}{{\cal D}^{\alpha}{\cal D}_{\alpha}}
\newcommand{\DbDb}{{\cal D}_{\dot{\alpha}}{\cal D}^{\dot{\alpha}}}

\newcommand{\hs}{\hspace{0.2cm}}
\newcommand{\dg}{g_{_{(1)}}}
\newcommand{\dgg}{g_{_{(2)}}}
\newcommand{\dgm}{g_{_{(m)}}}
\newcommand{\df}{f_{_{(1)}}}

\newcommand{\dfm}{f_{_{(m)}}}
\newcommand{\dilaton}{{\ell}}

\begin{document}
\begin{titlepage}
\begin{center}
\today\     \hfill    LBNL-38590 \\
            \hfill    UCB-PTH-96/13 \\
            \hfill LPTHE-Orsay 96/32 \\

{\large \bf Dilaton Stabilization in the Context \\
            of Dynamical Supersymmetry Breaking 
         \\ through Gaugino Condensation }
\footnote{This work was supported in part by the Director, Office of 
Energy Research, Office of High Energy and Nuclear Physics, Division 
of High Energy Physics of the U.S. Department of Energy under 
Contract DE-AC03-76SF00098 and in part by the National Science 
Foundation under grant PHY-95-14797.}\\[.1in]

    Pierre Bin\'{e}truy \\[.05in]

{\em  Laboratoire de Physique Th\'{e}orique et Hautes Energies,\footnote{
Laboratoire associ\'e au CNRS--URA-D0063.}\\ 
      Universit\'{e} Paris-Sud, F-91405 Orsay, France}\\[.05in]          

Mary K. Gaillard and Yi-Yen Wu \\[.05in]

{\em  Department of Physics and Theoretical Physics Group, 
      Lawrence Berkeley Laboratory, University of California,
      Berkeley, CA 94720, USA}\\[.2in] 
\end{center}

\begin{abstract}
      We study gaugino condensation in the context of superstring effective 
theories using the linear multiplet formulation for the dilaton superfield.
Including nonperturbative corrections to the K\"{a}hler potential 
for the dilaton may naturally achieve dilaton stabilization, with 
supersymmetry breaking and gaugino condensation; these three issues are 
interrelated in a very simple way. In a toy model with a single static
condensate, a dilaton $vev$ is found within a phenomenologically interesting 
range. The effective theory differs significantly from condensate models 
studied previously in the chiral formulation.

\end{abstract}
\end{titlepage}
\renewcommand{\thepage}{\roman{page}}
\setcounter{page}{2}
\mbox{ }

\vskip 1in

\begin{center}
{\bf Disclaimer}
\end{center}

\vskip .2in

\begin{scriptsize}
\begin{quotation}
This document was prepared as an account of work sponsored by the United
States Government.  Neither the United States Government nor any agency
thereof, nor The Regents of the University of California, nor any of their
employees, makes any warranty, express or implied, or assumes any legal
liability or responsibility for the accuracy, completeness, or usefulness
of any information, apparatus, product, or process disclosed, or represents
that its use would not infringe privately owned rights.  Reference herein
to any specific commercial products process, or service by its trade name,
trademark, manufacturer, or otherwise, does not necessarily constitute or
imply its endorsement, recommendation, or favoring by the United States
Government or any agency thereof, or The Regents of the University of
California.  The views and opinions of authors expressed herein do not
necessarily state or reflect those of the United States Government or any
agency thereof of The Regents of the University of California and shall
not be used for advertising or product endorsement purposes.
\end{quotation}
\end{scriptsize}

\vskip 2in

\begin{center}
\begin{small}
{\it Lawrence Berkeley Laboratory is an equal opportunity employer.}
\end{small}
\end{center}

\newpage
\renewcommand{\thepage}{\arabic{page}}
\renewcommand{\theequation}{\arabic{section}.\arabic{equation}}
\setcounter{page}{1}
\section{Introduction}
\hspace{0.8cm}\setcounter{equation}{0}
Among the massless string modes, a real scalar (dilaton), 
an antisymmetric tensor field (the Kalb-Ramond field) and 
their supersymmetric partners can be described either by a 
chiral superfield $S$ or by a linear multiplet $L$, which is 
known as the chiral-linear duality. By definition, the linear 
multiplet $L$ is a vector superfield that satisfies the 
following constraints \cite{Linear}:
\bea
-(\DbDb-8R)L\,&=&\,0, \nonumber\\
-(\DaDa-8R^{\dagger})L\,&=&\,0.
\eea
The lowest component of $L$ is the dilaton field 
$\dilaton$, and its $vev$ is related to the gauge 
coupling constant as follows:
$g^{2}(M_{S})\,=\,2\langle\dilaton\rangle$,
where $M_{S}$ is the string scale \cite{Gaillard92,KL}.
Although the chiral-linear duality is obvious at tree level, 
it becomes obscure when quantum effects are included. 
Although scalar-2-form field strength duality, which is contained 
in chiral-linear duality, has been shown to be preserved in perturbation 
theory~\cite{frad}, the situation is less clear in the presence of 
nonperturbative effects, which are important in the study of gaugino 
condensation. It has recently been shown~\cite{Burgess95,Binetruy95} that
gaugino condensation can be formulated directly using a linear multiplet
for the dilaton. However, the content of the resulting chiral-linear duality 
transformation is in general very complicated. If there is an elegant 
description of gaugino condensates in the context of superstring effective 
theories, it may be simple in only one of these formulations, but not 
in both. Therefore, a pertinent issue is: which formulation is better?

In this paper we will construct the effective 
theory of gaugino condensation directly in the linear multiplet 
formulation without referring to the chiral formulation.
There is reason to believe that the linear multiplet formulation 
is in fact more appropriate. 
The stringy reason for choosing the linear multiplet formulation is 
that the precise field content of the linear multiplet appears in the 
massless string spectrum, and $\langle L\rangle$ 
plays the role of string loop expansion parameter. Therefore, 
string information is more naturally encoded in the linear 
multiplet formulation of string effective theory. In the context
of gaugino condensation, it has been pointed out that the
gaugino condensate $U$ should be a constrained chiral superfield 
\cite{Burgess95,Binetruy95,Pillon}; this constraint arises naturally
in the linear multiplet formulation of gaugino condensation. Finally, 
in the linear formulation the symmetries of the underlying 
Yang-Mills theory in the weak coupling limit are automatically
respected~\cite{sduality}. 

In the next section we describe the linear multiplet formulation of
string effective Yang-Mills theory, whose effective theory below
the condensation scale is constructed and analyzed in Sect. 3. 
It is then shown in Sect. 4 that supersymmetry is broken and the dilaton
is stabilized in a large class of models of gaugino condensation. 
In this paper we use the K\"{a}hler superspace
formulation~\cite{Binetruy90}, suitably extended to incorporate the
linear multiplet~\cite{Binetruy91}.
\section{The Linear Multiplet Formulation}
\setcounter{equation}{0} 
\subsection{Superstring Effective Yang-Mills Theory}
\hspace{0.8cm}
In the realm of superstring effective Yang-Mills theory, 
there are two important ingredients, namely, the symmetry
group of modular transformations and the linear multiplet.
In order to make the discussion as explicit as possible, 
we consider here orbifolds with gauge group 
$\mbox{E}_{8}\otimes\mbox{E}_{6}\otimes\mbox{U(1)}^{2}$,
which have been studied most extensively in the context 
of modular symmetries \cite{Gaillard92,KL,Dixon90}. They contain three 
untwisted (1,1) moduli $T^{I}$, $I=1,\,2,\,3$, which transform 
under SL(2,Z) as follows:
\be
T^{I}\;\rightarrow\;\frac{aT^{I}-ib}{icT^{I}+d},\;\;\;
ad-bc=1,\;\;\;a,b,c,d\;\in\mbox{Z}.
\ee
The corresponding K\"{a}hler potential is
\be
G\,=\,\sum_{I}g^{I}\,+\,
      \sum_{A}\exp(\sum_{I}q_{A}^{I}g^{I})|\Phi^{A}|^{2}\,+\,
      \mbox{O}(\Phi^{4}),  
\ee
where $g^{I}\,=\,-\ln(T^{I}+\bar{T}^{I})$, and the modular weights
$q_{A}^{I}$ depend on the particular matter field $\Phi^{A}$
as well as on the modulus $T^{I}$. However, it is well known 
that the effective theory obtained from the massless truncation 
of superstring is {\it not} invariant under the modular 
transformations (2.1) at one loop \cite{Derendinger92,Ovrut93}.
Counterterms, that correspond to the result of 
integrating out massive modes, have to be added to the 
effective theory in order to restore modular invariance 
since string theory is known to be modular invariant to all 
orders of the loop expansion \cite{Giveon89}. Two types of such
counterterms have been discussed in the literature 
\cite{Gaillard92,Dixon90,Ovrut93}, the so-called $f$-type counterterm
and the Green-Schwarz counterterm. The Green-Schwarz counterterm, 
which is analogous to the Green-Schwarz anomaly cancellation 
mechanism in D=10, is naturally implemented with the linear multiplet 
formulation~\cite{Linear}. Here we consider only those orbifolds for 
which the full modular anomaly is cancelled by the Green-Schwarz 
counterterm alone. This is the case unless the 
modulus $T^{I}$ corresponds to an internal plane which is 
left invariant under some orbifold group transformations, 
which may happen only if an $N$=2 supersymmetric twisted 
sector is present~\cite{ant}. Therefore, a large class of orbifolds, 
including the $\mbox{Z}_{3}$ and $\mbox{Z}_{7}$ orbifolds, 
is under consideration here.

The antisymmetric tensor field of superstring theories undergoes 
Yang-Mills gauge transformations. In the effective theory, it can
be incorporated into a gauge invariant vector superfield $L$, the 
so-called modified linear multiplet, coupled to the Yang-Mills 
degrees of freedom as follows:
\bea
-(\DbDb-8R)L\,&=&\,(\DbDb-8R)\Omega\,=\,\sum_{a}\WaWa^{a}, 
\nonumber \\ 
-(\DaDa-8R^{\dagger})L\,&=&\,(\DaDa-8R^{\dagger})\Omega\,=\,
\sum_{a}\WbWb^{a},
\eea
where $\Omega$ is the Yang-Mills Chern-Simons superform. The 
summation extends over the indices $a$ numbering simple subgroups
of the full gauge group. The modified linear multiplet $L$ 
contains the linear multiplet as well as the Chern-Simons superform, 
and its gauge invariance is ensured by imposing appropriate 
transformation properties for the linear multiplet. The generic 
lagrangian describing the linear multiplet coupled to supergravity 
and matter in the presence of Yang-Mills Chern-Simons superform is
\cite{Gaillard92}:
\bea
K\,&=&\,k(L)\,+\,G, \nonumber \\
\Lag\,&=&\,-3\superint\,E\,F(L) \,+\,
\superint\,E\,\{\,bL\sum_{I}g^{I}\,\}, \\
b\,&=&\,\frac{\,C}{\,8\pi^{2}}\,=\,
        \frac{\,2\,}{\,3\,}b_{0},
\eea
where $L$ is the modified linear multiplet and $C\,=\,30$ 
is the Casimir operator in the adjoint representation of 
$\mbox{E}_{8}$. $b_{0}$ is the $\mbox{E}_{8}$ one-loop 
$\beta$-function coefficient. The first term of $\Lag$ 
is the superspace integral which yields the 
kinetic actions for the linear multiplet, supergravity, 
matter and Yang-Mills fields. The second term in (2.4) is 
the Green-Schwarz counterterm, which is ``minimal'' in the 
sense of \cite{Gaillard92}. Furthermore, arbitrariness in
the two functions $k(L)$ and $F(L)$ is reduced  by the
requirement that the Einstein term in $\Lag$ be canonical.
Under this constraint,  $k(L)$ and $F(L)$
are related to each other by the following first-order
differential equation~\cite{Binetruy91}: 
\be
F\,-\,L\frac{\diff F}{\diff L}\,=\,
1\,-\,\frac{1}{3}L\frac{\diff k}{\diff L}.
\ee
The complete component lagrangian of
(2.4) with the tree-level K\"{a}hler potential (i.e., $k(L)=\ln L$
and $F(L)=\frac{\,2\,}{\,3\,}$) has been presented in 
\cite{Adamietz93} based on the K\"{a}hler superspace formalism. 
Similar studies have also been performed in the 
superconformal formalism of supergravity \cite{Derendinger94}.
In the following sections, we are interested in the effective 
lagrangian of (2.4) below the condensation scale.
\subsection{The Low-Energy Effective Degrees of Freedom}
\hspace{0.8cm}
Below the condensation scale at which the gauge interaction becomes
strong, the effective lagrangian of the Yang-Mills sector can be 
described by a composite chiral superfield $U$, which corresponds 
to the chiral superfield $\WaWa$ of the underlying theory. (We 
consider here gaugino condensation of a simple gauge group.)
The scalar component of $U$ is naturally interpreted as the 
gaugino condensate. It was pointed out only recently that the composite 
field $U$ is actually a constrained chiral superfield 
\cite{Binetruy95}--\cite{sduality},\cite{Pillon}. The constraint on $U$ can be 
seen most clearly through the constrained superspace geometry of the 
underlying Yang-Mills theory. As a consequence of this constrained 
geometry, the chiral superfield $\WaWa$ and its hermitian conjugate 
$\WbWb$ satisfy the following constraint:
\be
(\DaDa-24R^{\dagger})\WaWa\,-\,(\DbDb-24R)\WbWb
\,=\,\mbox{total derivative.}
\ee
(2.7) has a natural interpretation in the context of a 3-form
supermultiplet, and indeed $\WaWa$ can be interpreted as the degrees 
of freedom of the 3-form field strength \cite{3form}. The explicit 
solution to the constraint (2.7) has been presented in \cite{Pillon},
and it allows us to identify the constrained chiral superfield $\WaWa$
with the chiral projection of an unconstrained vector superfield $L$:
\bea
\WaWa\,&=&\,-(\DbDb-8R)L, \nonumber \\ 
\WbWb\,&=&\,-(\DaDa-8R^{\dagger})L.
\eea
Below the condensation scale, the constraint (2.7) is replaced by the 
following constraint on $U$ and $\bar{U}$:
\be
(\DaDa-24R^{\dagger})U\,-\,(\DbDb-24R)\bar{U}
\,=\,\mbox{total derivative.} 
\ee
Similarly, the solution to (2.9) allows us to identify the constrained
chiral superfield $U$ with the chiral projection of an unconstrained
vector superfield $V$:
\bea
U\,&=&\,-(\DbDb-8R)V, \nonumber \\ 
\bar{U}\,&=&\,-(\DaDa-8R^{\dagger})V.
\eea
(2.10) is the explicit constraint on $U$ and $\bar{U}$.

In fact, the constraint on $U$ and $\bar{U}$ enters the linear 
multiplet formulation of gaugino condensation very naturally. As
described in Sect. 2.1, the linear multiplet formulation of 
supersymmetric Yang-Mills theory is described by a gauge-invariant 
vector superfield $L$ which satisfies
\bea
-(\DbDb-8R)L\,&=&\,(\DbDb-8R)\Omega\,=\,\WaWa, \nonumber \\ 
-(\DaDa-8R^{\dagger})L\,&=&\,(\DaDa-8R^{\dagger})\Omega\,=\,
\WbWb.
\eea
For the linear multiplet formulation of the effective lagrangian below
the condensation scale, (2.11) is replaced by
\bea
-(\DbDb-8R)V\,&=&\,U,\nonumber \\ 
-(\DaDa-8R^{\dagger})V\,&=&\,\bar{U},
\eea
where $U$ is the gaugino condensate chiral superfield, and $V$ 
contains the linear multiplet as well as the ``fossil'' Chern-Simons 
superform. In view of (2.12), it is clear that the constraint on $U$
and $\bar{U}$ arises naturally in the linear multiplet formulation 
of gaugino condensation. Furthermore, the low-energy degrees of 
freedom (i.e., the linear multiplet and the gaugino condensate) are 
nicely merged into a single vector superfield $V$, and therefore the 
linear multiplet formulation of gaugino condensation can elegantly be
described by $V$ alone. The detailed construction of the effective 
lagrangian for the vector superfield $V$ will be presented in the next 
section.
\section{Gaugino Condensation in Superstring \newline Effective Theory} 
\setcounter{equation}{0}
\subsection{A Simple Model}
\hspace{0.8cm}
Constructing the linear multiplet formulation of gaugino condensation
requires the specification of two functions of the vector superfield $V$, 
namely, the superpotential and the K\"{a}hler potential. In the linear 
multiplet formulation, there is no classical superpotential~\cite{sduality}, 
and the quantum superpotential originates from the nonperturbative effects 
of gaugino condensation. This nonperturbative superpotential, whose 
form was dictated by the anomaly structure of the underlying theory, 
was first obtained by Veneziano and Yankielowicz \cite{Veneziano82}. 
The details of its generalization to the case of matter coupled to $N$=1 
supergravity in the K\"{a}hler superspace formalism has been presented 
in \cite{chiral91}, and the superpotential term in the Lagrangian 
reads:
\bea
\superint\,\frac{E}{R}\,e^{K/2}W_{VY}\,&=&\,
\superint\,\frac{E}{R}\,\frac{1}{8}bU\ln(e^{-K/2}U/\mu^{3}),
\nonumber\\
\superint\,\frac{E}{R^{\dagger}}\,e^{K/2}\bar{W}_{VY}\,&=&\,
\superint\,\frac{E}{R^{\dagger}}
\,\frac{1}{8}b\bar{U}\ln(e^{-K/2}\bar{U}/\mu^{3}),
\eea
where $\,U\,=\,-(\DbDb-8R)V\,$ is the 
constrained gaugino condensate chiral superfield with K\"{a}hler
weight 2, and $\mu$ is a constant with dimension of mass that
is left undetermined by the method of anomaly matching. 

As for the K\"{a}hler potential for $V$, there is little knowledge 
beyond tree level. The best we can do at present is to 
treat all physically reasonable K\"{a}hler potentials on the same 
footing and to look for possible general features and/or interesting 
special cases. Before discussing this general analysis, it is 
instructive to examine a simple linear multiplet model for gaugino 
condensation defined as follows~\cite{sduality}:
\bea
K\,&=&\,\ln V\,+\,G, \nonumber \\
\Lag_{eff}\,&=&\,\superint\,E\,\{\,-2\,+\,bVG\,\} \,+\, 
\superint\,\frac{E}{R}\,e^{K/2}W_{VY} \,+\,
\superint\,\frac{E}{R^{\dagger}}\,e^{K/2}\bar{W}_{VY},
\nonumber \\
G\,&=&\,-\sum_{I}\ln(T^{I}+\bar{T}^{I}).
\eea
This simple model describes the effective theory for (2.4) below the
condensation scale, where the K\"{a}hler potential of $V$ assumes its
tree-level form. It is a ``static'' model in the sense that
no kinetic term for $U$ is included. From the viewpoint of the anomaly 
structure, static as well as nonstatic models are interesting 
in their own right. In the chiral formulation of gaugino condensation, 
it can be shown that the static model corresponds to the effective 
theory of the nonstatic model after the gaugino condensate $U$ is 
integrated out. Nonstatic models~\cite{Burgess95,Binetruy95} in the linear 
multiplet formulation have been studied less extensively. Here we will restrict
our attention to the static case, since the points we wish to illustrate are
not substantially altered by including a kinetic term for $U$. In Sect. 5 we 
will indicate how the model considered here can be generalized to the case of a
dynamical condensate.

With $\,U\,=\,-(\DbDb-8R)V\,$ and 
$\,\bar{U}\,=\,-(\DaDa-8R^{\dagger})V\,$, we can rewrite the 
superpotential terms of $\Lag_{eff}$ as a single D-term, and  
therefore the simple model (3.2) can be rewritten as follows:
\bea
K\,&=&\,\ln V\,+\,G, \nonumber \\
\Lag_{eff}\,&=&\,\superint\,E\,\{\,-2 \,+\, bVG \,+\,
bV\ln(e^{-K}\bar{U}U/\mu^{6})\,\}.
\eea
In (3.3), the modular anomaly cancellation by the Green-Schwarz 
counterterm is transparent~\cite{sduality}. The Green-Schwarz counterterm 
$\,bVG\,$ 
and the superpotential D-term $\,bV\ln(e^{-K}\bar{U}U/\mu^{6})\,$ are
{\em not} modular invariant separately, but their sum is modular 
invariant, which ensures the modular invariance of the full theory. 
In fact, the Green-Schwarz counterterm cancels the $T^{I}$ 
moduli-dependence of the superpotential completely. This is a unique 
feature of the linear multiplet formulation, and, as we will see later, 
has interesting implications for the moduli-dependence of physical 
quantities. 

Throughout this paper only the bosonic and gravitino parts of the component 
lagrangian are presented, since we are interested in the vacuum configuration
and the gravitino mass. In the following, we enumerate the definitions of 
bosonic component fields of the vector superfield $V$.
\bea
\dilaton\,&=&\,V\lowest,\nonumber\\
\sigma^{m}_{\alpha\dot{\alpha}}B_{m}\,&=&\,
\frac{1}{2}[\,\Diff_{\alpha},\Diff_{\dot{\alpha}}\,]V\lowest\,+\,
\frac{2}{3}\dilaton\sigma^{a}_{\alpha\dot{\alpha}} b_{a},\nonumber\\
u\,&=&\,U\lowest\,=\,-(\bar{\Diff}^{2}-8R)V\lowest,\nonumber\\
\bar{u}\,&=&\,\bar{U}\lowest\,=\,-(\Diff^{2}-8R^{\dagger})V\lowest,
\nonumber \\
D\,&=&\,\frac{1}{8}\Diff^{\beta}(\bar{\Diff}^{2}-8R)
      \Diff_{\beta}V\lowest\nonumber\\
   &=&\,\frac{1}{8}\Diff_{\dot{\beta}}(\Diff^{2}-8R^{\dagger})
      \Diff^{\dot{\beta}}V\lowest,
\eea
where 
\be
-\,\frac{1}{6}M\,=\,R\lowest,\;\;
-\,\frac{1}{6}\bar{M}\,=\,R^{\dagger}\lowest,\;\;
-\,\frac{1}{3}b_{a}\,=\,G_{a}\lowest
\ee
are the auxiliary components of supergravity multiplet. It is 
convenient to write the lowest components of $\Diff^{2}U$ and 
$\bar{\Diff}^{2}\bar{U}$ as follows:
\be
-4F_{U}\,=\,\Diff^{2}U\lowest, \;\;\; 
-4\bar{F}_{\bar{U}}\,=\,\bar{\Diff}^{2}\bar{U}\lowest.
\ee
$(F_{U}-\bar{F}_{\bar{U}})$ can be explicitly expressed as follows:
\be
(F_{U}-\bar{F}_{\bar{U}})\,=\,4i\nabla^{m}\!B_{m}
\,+\,u\bar{M}\,-\,\bar{u}M.
\ee
The expression for $(F_{U}+\bar{F}_{\bar{U}})$ contains the auxiliary 
field $D$. The bosonic components of $T^{I}$ and $\bar{T}^{I}$ are
\bea
t^{I}\,&=&\,T^{I}\lowest,\;\;\;
-4F_{T}^{I}\,=\,\Diff^{2}T^{I}\lowest,\nonumber\\
\bar{t}^{I}\,&=&\,\bar{T}^{I}\lowest,\;\;\;
-4\bar{F}_{\bar{T}}^{I}\,=\,\bar{\Diff}^{2}\bar{T}^{I}\lowest.
\eea
We leave the details of constructing the component lagrangian for
this simple model (in the K\"{a}hler superspace formalism) to 
Sect. 3.2, and present here only the scalar potential:
\be
V_{pot}\,=\,\frac{1}{16e^{2}\dilaton}
(\,1\,+\,2b\dilaton\,-\,2b^{2}\dilaton^{2}\,)\mu^{6}e^{-\,1/{b\dilaton}}.
\ee
Eq.(3.9) agrees with the result obtained in~\cite{Binetruy95}, where the 
model defined by (3.2) was studied for the case of a single modulus using the 
superconformal formalism of supergravity.

However, this simple model is not viable. As expected, the 
weak-coupling limit $\dilaton=0$ is always a minimum. As shown in Fig.1, 
the scalar potential starts with $V_{pot}=0$ at $\dilaton=0$, first 
rises and then falls without limit as $\dilaton$ increases. Therefore, 
$V_{pot}$ is unbounded from below, and this simple model has no 
well-defined vacuum. This may be somewhat surprising because the model 
defined by (3.2) superficially appears to be of the no-scale type: 
the Green-Schwarz counterterm, that destroys the no-scale property of chiral 
models and destabilizes the potential, is cancelled here by quantum effects 
that induce a potential for the condensate.  However the resulting quantum 
contribution to the Lagrangian (3.3), $bV\ln(U\bar{U}/V)$, has an implicit 
$T^I$-dependence through the superfield $U$ due to its nonvanishing K\"ahler 
weight: $w(U)= 2$. This implicit moduli-dependence is a consequence of the 
anomaly matching condition, and parallels the construction of the effective 
theory in the chiral formulation~\cite{Veneziano82} which is also not of the 
no-scale form once the Green-Schwarz counterterm is included.  By contrast, 
in~\cite{sduality} a no-scale model was constructed in the chiral formulation 
precisely through a cancellation of the Green-Schwarz counterterm. In the 
construction of that model, the point of view was adopted that a superpotential 
for the dilaton could arise only from nonperturbative effects on the string 
world sheet, and the anomaly matching condition was bypassed by directly 
writing an effective low energy theory that was exactly modular invariant. 
The relation between these approaches warrants further investigation.

If we take a closer look at (3.9), it is clear that the unboundedness of 
$V_{pot}$ in the strong-coupling limit $\:\dilaton\,\rightarrow\,\infty\:$
is caused by a term of two-loop order: $\:-2b^{2}\dilaton^{2}\:$.  This 
observation strongly suggests that the underlying reason for unboundedness
is our poor control over the model in the strong-coupling regime. 
The form of the superpotential $W_{VY}$ is completely fixed by the underlying 
anomaly structure.  However the K\"{a}hler potential is much less constrained,
and the choice (3.2) cannot be expected to be valid in the strong-coupling 
regime where the nonperturbative contributions should not be ignored. We
conclude that the unboundedness shown in Fig. 1 simply simply reflects the 
importance of nonperturbative contributions~\cite{Derendinger95,Shenker90} 
to the K\"{a}hler potential. In the absence of a better knowledge of the exact 
K\"{a}hler potential, we will consider models with generic K\"{a}hler 
potentials in the following sections.
\subsection{General Static Model}
\hspace{0.8cm}
In this section, we show how to construct the component lagrangian
for generic linear multiplet models of gaugino condensation in the 
K\"{a}hler superspace formalism. Further computational details can be 
found in \cite{Binetruy90,Adamietz93}.  Although our results can 
probably be rephrased in the chiral formulation, the equivalent chiral
superfield formulation may be expected to be rather complicated because of
the constraint on the condensate chiral superfield $U$.  Quite generally
we do not expect a simple ansatz in one formalism to appear simple in the
other.

As suggested in Sect. 3.1, 
we extend the simple model in (3.2) to linear multiplet models of 
gaugino condensation with generic K\"{a}hler potentials defined 
as follows: 
\bea
K\,&=&\,\ln V\,+\,g(V)\,+\,G, \nonumber \\
\Lag_{eff}\,&=&\,\superint\,E\,\{\,(\,-2\,+\,f(V)\,)\,+\, 
bVG\,+\,bV\ln(e^{-K}\bar{U}U/\mu^{6})\,\}.\hspace{1.5cm}
\eea  
For convenience, we also write $\;\ln V\,+\,g(V)\,\equiv\,k(V).\;$ 
$g(V)$ and $f(V)$ represent quantum corrections to the tree-level 
K\"{a}hler potential, and, according to (2.6), they are unambiguously 
related to each other by the following first-order differential 
equation:
\be
V\frac{\diff g(V)}{\diff V}\,=\,
-V\frac{\diff f(V)}{\diff V}\,+\,f,
\ee
\be
g(V=0)\,=\,0 \;\;\;\mbox{and}\;\;\; f(V=0)\,=\,0. 
\ee
The boundary condition of $g(V)$ and $f(V)$ at $V=0$ (the 
weak-coupling limit) is fixed by the tree-level K\"{a}hler
potential. Before trying to specify $g(V)$ and $f(V)$, it is 
reasonable to assume for the present that $g(V)$ and $f(V)$ are 
arbitrary but bounded.

In the construction of the component field lagrangian, we use the
chiral density multiplet method~\cite{Binetruy90}, which provides us 
with the locally supersymmetric generalization of the F-term 
construction in global supersymmetry. The chiral density multiplet 
${\bf r}$ and its hermitian conjugate ${\bf \bar{r}}$ for the generic 
model in (3.10) are: 
\bea
{\bf r}\,&=&\,-\,\frac{1}{8}(\bar{\Diff}^{2}-8R)\{\,(\,-2\,+\,f(V)\,)
\,+\,bVG\,+\,bV\ln(e^{-K}\bar{U}U/\mu^{6})\,\}, \nonumber\\
{\bf \bar{r}}\,&=&\,-\,\frac{1}{8}(\Diff^{2}-8R^{\dagger})\{\,(\,-2\,
+\,f(V)\,)\,+\,bVG\,+\,bV\ln(e^{-K}\bar{U}U/\mu^{6})\,\}.
\hspace{1.5cm}
\eea
In order to obtain the component lagrangian $\Lag_{eff}$, we need 
to work out the following expression
\bea
\frac{1}{e}\Lag_{eff}\,&=&\,-\,\frac{1}{4}\Diff^{2}{\bf r}
\lowest\,+\,\frac{i}{2}(\bar{\psi}_{m}\bar{\sigma}^{m})^{\alpha}
\Diff_{\alpha}{\bf r}\lowest\nonumber\\
& &\,-\,(\bar{\psi}_{m}\bar{\sigma}^{mn}\bar{\psi}_{n}+\bar{M})
{\bf r}\lowest \,\,+\,\, \mbox{h.c.}
\eea
An important point in the computation of (3.14) is the evaluation 
of the component field content of the K\"{a}hler supercovariant
derivatives, a rather tricky process. The details of this computation
have by now become general wisdom and we can to a large extent rely 
on the existing literature \cite{Wess}. In particular, the Lorentz
transformation and the K\"{a}hler transformation are incorporated in 
a very similar way in the K\"{a}hler superspace formalism, and the
Lorentz connection as well as the so-called K\"{a}hler connection 
$A_{M}$ are incorporated into the K\"{a}hler supercovariant 
derivatives in a concise and constructive way. The K\"{a}hler
connection $A_{M}$ is not an independent field but rather expressed
in terms of the K\"{a}hler potential $K$ as follows
\be
A_{\alpha}\,=\,\frac{1}{4}E_{\alpha}^{\hs M}\partial_{M}K,\;\;\;
A_{\dot{\alpha}}\,=\,-\,\frac{1}{4}
E_{\dot{\alpha}}^{\hs M}\partial_{M}K,
\ee
\be
\sigma^{a}_{\alpha\dot{\alpha}}A_{a}\,=\,
\frac{3}{2}i\sigma^{a}_{\alpha\dot{\alpha}}G_{a}\,-\,
\frac{1}{8}i[\,\Diff_{\alpha},\Diff_{\dot{\alpha}}\,]K.
\ee
In order to extract the explicit form of the various couplings, we  
choose to write out explicitly the vectorial part of the K\"{a}hler
connection and keep only the Lorentz connection in the definition of
covariant derivatives when we present the component expressions. In 
the following, we give the lowest component of the vectorial part of 
the K\"{a}hler connection $\: A_{m}\lowest\:$ for our generic model.
\be
A_{m}\,=\,e_{m}^{\hs a}A_{a}\,+\,
\frac{1}{2}\psi_{m}^{\hs\alpha}A_{\alpha}\,+\,
\frac{1}{2}\bar{\psi}_{m\dot{\alpha}}A^{\dot{\alpha}}.
\ee

\bea
A_{m}\lowest\,&=&\,-\,\frac{i}{4\dilaton}(\dilaton\dg+1)B_{m}\,+\,
\frac{i}{6}(\dilaton\dg-2)e_{m}^{\hs a}b_{a}\nonumber\\
& &\,+\,\sum_{I}\frac{1}{4(t^{I}+\bar{t}^{I})}
(\nabla_{\!m}\bar{t}^{I}\,-\,\nabla_{\!m}t^{I}).
\eea

\bea
\dgm\,&=&\,\dgm\!(\dilaton)\,=\,
\frac{\diff^m g(V)}{\diff V^m}\lowest,\nonumber\\
\dfm\,&=&\,\dfm\!(\dilaton)\,=\,
\frac{\diff^m f(V)}{\diff V^m}\lowest.
\eea

Another hallmark of the K\"{a}hler superspace formalism are the 
chiral superfield $X_{\alpha}$ and the antichiral superfield  
$\bar{X}^{\dot{\alpha}}$. They arise in complete analogy with usual
supersymmetric abelian gauge theory except that now the corresponding
vector superfield is replaced by the K\"{a}hler potential:
\bea
X_{\alpha}\,&=&\,-\,\frac{1}{8}(\DbDb-8R)
\Diff_{\alpha}K,\nonumber\\
\bar{X}^{\dot{\alpha}}\,&=&\,-\,\frac{1}{8}(\DaDa-8R^{\dagger})
\Diff^{\dot{\alpha}}\!K.
\eea
In the computation of (3.14), we need to decompose the lowest 
components of the following six superfields:
$X_{\alpha}$, $\bar{X}^{\dot{\alpha}}$, $\Diff_{\alpha}R$,
$\Diff^{\dot{\alpha}}\!R^{\dagger}$, 
$(\Diff^{\alpha}\!X_{\alpha}+\Diff_{\dot{\alpha}}\bar{X}^{\dot{\alpha}})$
and $(\Diff^{2}\!R+\bar{\Diff}^{2}\!R^{\dagger})$ into component 
fields. This is done by solving the following six simple algebraic 
equations:
\bea
(V\frac{\diff g}{\diff V}+1)\Diff_{\alpha}R\,+\,
X_{\alpha}\,&=&\,\Xi_{\alpha},\\
3\Diff_{\alpha}R\,+\,X_{\alpha}\,&=&\,
-2(\sigma^{cb}\epsilon)_{\alpha\varphi}T_{cb}^
{\hs\hspace{0.04cm}\varphi}.
\eea

\bea
(V\frac{\diff g}{\diff V}+1)\Diff^{\dot{\alpha}}\!R^{\dagger}
\,+\,\bar{X}^{\dot{\alpha}}\,&=&\,\bar{\Xi}^{\dot{\alpha}},\\
3\Diff^{\dot{\alpha}}\!R^{\dagger}\,+\,\bar{X}^{\dot{\alpha}}\,&=&\,
-2(\bar{\sigma}^{cb}\epsilon)^{
\dot{\alpha}\dot{\varphi}}T_{cb\dot{\varphi}}.
\eea

\bea
(V\frac{\diff g}{\diff V}+1)(\Diff^{2}\!R+\bar{\Diff}^{2}\!
R^{\dagger})\,+\,(\Diff^{\alpha}\!X_{\alpha}+
\Diff_{\dot{\alpha}}\bar{X}^{\dot{\alpha}})\,&=&\,\Delta,\\
3(\Diff^{2}\!R+\bar{\Diff}^{2}\!R^{\dagger})\,+\,(\Diff^{\alpha}
\!X_{\alpha}+\Diff_{\dot{\alpha}}\bar{X}^{\dot{\alpha}})\,&=&\,
-2R_{ba}^{\hs ba}\,+\,12G^{a}G_{a} \hspace{0cm}\nonumber\\
\,& &\,+\,96RR^{\dagger}. \hspace{0cm}
\eea
The identities (3.22), (3.24) and (3.26) arise solely from the 
structure of K\"{a}hler superspace. (3.22) and (3.24) involve the
torsion superfields $T_{cb}^{\hs\hspace{0.04cm}\varphi}$ and 
$T_{cb\dot{\varphi}}$, which in their lowest components contain 
the curl of the Rarita-Schwinger field. The identities (3.21), 
(3.23) and (3.25) arise directly from the definitions of 
$X_{\alpha}$, $\bar{X}^{\dot{\alpha}}$, 
$(\Diff^{\alpha}\!X_{\alpha}+\Diff_{\dot{\alpha}}\bar{X}^{\dot{\alpha}})$,
and therefore they depend on the K\"{a}hler potential explicitly. 
Computing $X_{\alpha}$, $\bar{X}^{\dot{\alpha}}$ and  
$(\Diff^{\alpha}\!X_{\alpha}+\Diff_{\dot{\alpha}}\bar{X}^{\dot{\alpha}})$
according to (3.20) defines the contents of $\Xi_{\alpha}$,
$\bar{\Xi}^{\dot{\alpha}}$ and $\Delta$ respectively. In the 
following, we present the component field expressions of the lowest
components of $\Xi_{\alpha}$, $\bar{\Xi}^{\dot{\alpha}}$ and $\Delta$.
\bea
& &\,\frac{i}{2}(\bar{\psi}_{m}\bar{\sigma}^{m})^{\alpha}
\Xi_{\alpha}\lowest\,-\,\frac{i}{2}\bar{\Xi}_{\dot{\alpha}}
(\bar{\sigma}^{m}\psi_{m})^{\dot{\alpha}}\lowest \nonumber\\
&=&\,-\,\frac{1}{8\dilaton}(\dilaton\dg+1)(\,\bar{u}\,+\,
\frac{4}{3}\dilaton\bar{M}\,)(\psi_{m}\sigma^{mn}\psi_{n})
\nonumber\\
& &\,-\,\frac{1}{8\dilaton}(\dilaton\dg+1)(\,u\,+\,\frac{4}{3}\dilaton M\,)
(\bar{\psi}_{m}\bar{\sigma}^{mn}\bar{\psi}_{n})
\nonumber\\
& &\,+\,\frac{i}{4\dilaton}(\dilaton\dg+1)
(\,\eta^{mn}\eta^{pq}\,-\,\eta^{mq}\eta^{np}\,)
(\bar{\psi}_{m}\bar{\sigma}_{n}\psi_{p})\,\nabla_{\!q}\dilaton
\nonumber\\
& &\,+\,\frac{i}{6}(\dilaton\dg+1)\epsilon^{mnpq}
(\bar{\psi}_{m}\bar{\sigma}_{n}\psi_{p})
e_{q}^{\hspace{0.12cm} a}b_{a}
\nonumber\\
& &\,-\,\frac{i}{4\dilaton}(\dilaton\dg+1)\epsilon^{mnpq}
(\bar{\psi}_{m}\bar{\sigma}_{n}\psi_{p})B_{\!q}
\nonumber\\
& &\,-\,\frac{1}{4}(\Diff^{a}\Diff^{\alpha}k)
\psi_{a\alpha}\lowest
\,-\,\frac{1}{4}\bar{\psi}_{a\dot{\alpha}}
(\Diff^{a}\Diff^{\dot{\alpha}}k)\lowest.
\eea
The way $\Xi_{\alpha}\lowest$ and $\bar{\Xi}^{\dot{\alpha}}\lowest$
are presented in (3.27) will be useful for the computation of (3.14).
\bea
& &\,\Delta\lowest \nonumber\\
&=&\,-\,\frac{1}{\dilaton^{2}}(\dilaton^{2}\!\dgg-1)
\nabla^{m}\!\dilaton\,\nabla_{\!m}\!\dilaton 
\,+\,\frac{1}{\dilaton^{2}}(\dilaton^{2}\!\dgg-1)B^{m}\!B_{m}
\nonumber\\
& &\,+\,4\sum_{I}\frac{1}{(t^{I}+\bar{t}^{I})^{2}}
\nabla^{m}\bar{t}^{I}\,\nabla_{\!m}t^{I}
\,-\,\frac{4}{9}(\dilaton^{2}\!\dgg-\dilaton\dg-2)\bar{M}\!M
\nonumber\\
& &\,+\,\frac{4}{9}(\dilaton^{2}\!\dgg+2\dilaton\dg+1)b^{a}b_{a}
\,-\,4\sum_{I}\frac{1}{(t^{I}+\bar{t}^{I})^{2}}
\bar{F}_{\bar{T}}^{I}F_{T}^{I}
\nonumber\\
& &\,-\,\frac{4}{3\dilaton}(\dilaton^{2}\!\dgg+\dilaton\dg)
B^{m}e_{m}^{\hs a}b_{a}
\,-\,\frac{1}{2\dilaton}(\dilaton\dg+1)(F_{U}+\bar{F}_{\bar{U}})
\nonumber\\
& &\,-\,\frac{1}{6\dilaton}(2\dilaton^{2}\!\dgg-\dilaton\dg-3)
(\,u\bar{M}\,+\,\bar{u}M\,)
\,-\,\frac{1}{4\dilaton^{2}}(\dilaton^{2}\!\dgg-1)\bar{u}u
\nonumber\\
& &\,+\,2\nabla^{m}\!\nabla_{\!m}k
\,-\,(\Diff^{a}\Diff^{\alpha}k)
\psi_{a\alpha}\lowest
\,-\,\bar{\psi}_{a\dot{\alpha}}
(\Diff^{a}\Diff^{\dot{\alpha}}k)\lowest.
\eea
It is unnecessary to decompose the last two terms in (3.27) and in 
(3.28) because they eventually cancel with one another. 

Eqs.(3.15-28) describe the key steps involved in the computation of 
(3.14). The rest of it is standard and will not be detailed here. In the
following, we present the component field expression of $\Lag_{eff}$ 
as the sum of the bosonic part $\Lag_{B}$ and the gravitino part 
$\Lag_{\tilde{G}}$ as follows.\footnote{Only the bosonic and gravitino
parts of the component field expressions are presented here.}
\be
\Lag_{eff}\,=\,\Lag_{B}\,+\,\Lag_{\tilde{G}}.
\ee

\bea
\frac{1}{e}\Lag_{B}\,&=&\,-\,\frac{1}{2}{\cal R}
\,-\,\frac{1}{4\dilaton^{2}}(\dilaton\dg+1)
\nabla^{m}\!\dilaton\,\nabla_{\!m}\!\dilaton
\nonumber\\
& &\,+\,\frac{1}{4\dilaton^{2}}(\dilaton\dg+1)B^{m}\!B_{m}
\,-\,(1+b\dilaton)\sum_{I}\frac{1}{(t^{I}+\bar{t}^{I})^{2}}
\nabla^{m}\bar{t}^{I}\,\nabla_{\!m}t^{I}
\nonumber\\
& &\,+\,\frac{1}{9}(\dilaton\dg-2)\bar{M}\!M
\,-\,\frac{1}{9}(\dilaton\dg-2)b^{a}b_{a}
\nonumber\\
& &\,+\,(1+b\dilaton)\sum_{I}\frac{1}{(t^{I}+\bar{t}^{I})^{2}}
\bar{F}_{\bar{T}}^{I}F_{T}^{I}
\nonumber\\
& &\,+\,\frac{1}{8\dilaton}\{\,f\,+\,1\,+\,b\dilaton\ln(e^{-k}\bar{u}u/
        \mu^{6})\,+\,2b\dilaton\,\}(F_{U}+\bar{F}_{\bar{U}})
\nonumber\\
& &\,-\,\frac{1}{8\dilaton}\{\,f\,+\,1\,+\,b\dilaton\ln(e^{-k}\bar{u}u/
\mu^{6})\,+\,\frac{2}{3}b\dilaton(\dilaton\dg+1)\,\}
(\,u\bar{M}\,+\,\bar{u}M\,)
\nonumber\\
& &\,-\,\frac{1}{16\dilaton^{2}}(1+2b\dilaton)(\dilaton\dg+1)\bar{u}u
\nonumber\\
& &\,-\,\frac{i}{2}b\ln(\frac{\bar{u}}{u})\nabla^{m}\!B_{m}
\,-\,\frac{i}{2}b\sum_{I}\frac{1}{(t^{I}+\bar{t}^{I})}
(\,\nabla^{m}\bar{t}^{I}\,-\,\nabla^{m}t^{I}\,)B_{m}.
\eea

\bea
\frac{1}{e}\Lag_{\tilde{G}}\,&=&\,\frac{1}{2}\epsilon^{mnpq}
(\,\bar{\psi}_{m}\bar{\sigma}_{n}\!\nabla_{\!p}\psi_{q}\,-\,
\psi_{m}\sigma_{n}\!\nabla_{\!p}\bar{\psi}_{q}\,)
\nonumber\\
& &\,-\,\frac{1}{8\dilaton}\{\,f\,+\,1\,+\,b\dilaton\ln(e^{-k}\bar{u}u/
\mu^{6})\,\}\,\bar{u}\,(\psi_{m}\sigma^{mn}\psi_{n})
\nonumber\\
& &\,-\,\frac{1}{8\dilaton}\{\,f\,+\,1\,+\,b\dilaton\ln(e^{-k}\bar{u}u/
\mu^{6})\,\}\,u\,(\bar{\psi}_{m}\bar{\sigma}^{mn}\bar{\psi}_{n})
\nonumber\\
& &\,-\,\frac{1}{4}(1+b\dilaton)\sum_{I}\frac{1}{(t^{I}+\bar{t}^{I})}
\epsilon^{mnpq}(\bar{\psi}_{m}\bar{\sigma}_{n}\psi_{p})
(\,\nabla_{\!q}\bar{t}^{I}\,-\,\nabla_{\!q}t^{I}\,)
\nonumber\\
& &\,+\,\frac{i}{4\dilaton}(1+b\dilaton)(\dilaton\dg+1)
(\,\eta^{mn}\eta^{pq}\,-\,\eta^{mq}\eta^{np}\,)
(\bar{\psi}_{m}\bar{\sigma}_{n}\psi_{p})\,\nabla_{\!q}\dilaton
\nonumber\\
& &\,-\,\frac{i}{4}b\dilaton
(\,\eta^{mn}\eta^{pq}\,-\,\eta^{mq}\eta^{np}\,)
(\bar{\psi}_{m}\bar{\sigma}_{n}\psi_{p})\,
\nabla_{\!q}\ln(\bar{u}u)
\nonumber\\
& &\,+\,\frac{1}{4}b\dilaton\,\epsilon^{mnpq}
(\bar{\psi}_{m}\bar{\sigma}_{n}\psi_{p})\,
\nabla_{\!q}\ln(\frac{\bar{u}}{u}).
\eea
For completeness, we also give the definitions of covariant derivatives:
\bea
\nabla_{\!m}\!\dilaton\,&=&\,\partial_{m}\!\dilaton,\;\;\;
\nabla_{\!m}t^{I}\,=\,\partial_{m}t^{I},\;\;\;
\nabla_{\!m}\bar{t}^{I}\,=\,\partial_{m}\bar{t}^{I},
\nonumber\\
\nabla_{\!m}\psi_{n}^{\hs\alpha}\,&=&\,
\partial_{m}\psi_{n}^{\hs\alpha}\,+\,
\psi_{n}^{\hs\beta}\omega_{m\beta}^{\hs\hs\alpha},
\;\;\;
\nabla_{\!m}\bar{\psi}_{n\dot{\alpha}}\,=\,
\partial_{m}\bar{\psi}_{n\dot{\alpha}}\,+\,
\bar{\psi}_{n\dot{\beta}}\,\omega_{m\hs\dot{\alpha}}
^{\hs\dot{\beta}}. \hspace{2cm}
\eea

To proceed further, we need to eliminate the auxiliary fields from
$\Lag_{eff}$ through their equations of motion. The equation of 
motion of the auxiliary field $(F_{U}+\bar{F}_{\bar{U}})$ is
\be
f\,+\,1\,+\,b\dilaton\ln(e^{-k}\bar{u}u/\mu^{6})\,+\,
2b\dilaton\,=\,0.
\ee
Eq. (3.33) implies that in static models the auxiliary field $\bar{u}u$
is expressed in terms of dilaton $\dilaton$. The equations of motion 
of $F_{T}^{I}$, $\bar{F}_{\bar{T}}^{I}$ and the auxiliary fields 
$b^{a}$, $M$, $\bar{M}$ of the supergravity multiplet are (if 
$\dilaton\dg-2 \neq 0$)
\bea
F_{T}^{I}\,&=&\,0, \;\;\;
\bar{F}_{\bar{T}}^{I}\,=\,0,\nonumber\\
b^{a}\,&=&\,0, \nonumber\\
M\,&=&\,\frac{\, 3\,}{\, 4\,}bu,\;\;\;
\bar{M}\,=\,\frac{\, 3\,}{\, 4\,}b\bar{u}.
\eea
Now we are left with only one auxiliary field to eliminate, where 
this auxiliary field can be either $\,\,i\ln({\bar{u}}/{u})$ or
$B_{m}$. This corresponds to the fact that there are two ways to
perform duality transformation. If we take 
$\,\,i\ln({\bar{u}}/{u})$ to be auxiliary, its equation of 
motion is
\be
\nabla_{\!q}\{\,B^{q}\,-\,\frac{i}{2}\dilaton\,\epsilon^{mnpq}
(\bar{\psi}_{m}\bar{\sigma}_{n}\psi_{p})\,\}\,=\,0,
\ee
which ensures that $\,\{B^{q}\,-\,\frac{i}{2}\dilaton\,\epsilon^{mnpq}
(\bar{\psi}_{m}\bar{\sigma}_{n}\psi_{p})\}\,$ is dual to the 
field strength of an antisymmetric tensor \cite{Binetruy95}.
The term $B^{m}\!B_{m}$ in the lagrangian $\Lag_{eff}$ thus 
generates a kinetic term of this antisymmetric tensor field and 
its coupling to the gravitino. The other way to perform the duality 
transformation is to treat $B_{m}$ as an auxiliary field by 
rewriting the term $\,-\,\frac{i}{2}b\ln({\bar{u}}/{u})
\nabla^{m}\!B_{m}$ in $\Lag_{eff}$ as $\,\frac{i}{2}bB^{m}
\nabla_{\!m}\!\ln({\bar{u}}/{u})$, and then to eliminate
$B_{m}$ from $\Lag_{eff}$ through its equation of motion as 
follows:
\bea
B_{m}\,&=&\,-\,i\frac{b\dilaton^{2}}{(\dilaton\dg+1)}\nabla_{\!m}\!
\ln(\frac{\bar{u}}{u}) \nonumber\\
& &\,+\,i\frac{b\dilaton^{2}}{(\dilaton\dg+1)}
\sum_{I}\frac{1}{(t^{I}+\bar{t}^{I})}
(\,\nabla_{\!m}\bar{t}^{I}\,-\,\nabla_{\!m}t^{I}\,).
\eea
The terms $B^{m}\!B_{m}$ and $\,\frac{i}{2}bB^{m}
\nabla_{\!m}\!\ln({\bar{u}}/{u})$ in $\Lag_{eff}$ 
will generate a kinetic term for $\,\,i\ln({\bar{u}}/{u})$. 
It is clear that $\,\,i\ln({\bar{u}}/{u})$ plays the 
role of the pseudoscalar dual to $B_{m}$ in the lagrangian obtained from the
above after a duality transformation.  With (3.33-36), it is
then trivial to eliminate the auxiliary fields from $\Lag_{eff}$.
The physics of $\Lag_{eff}$ will be investigated in the following sections.
\subsection{Gaugino Condensate and the Gravitino Mass}
\hspace{0.8cm}
Hidden-sector gaugino condensation in superstring effective 
theories is a very attractive scheme~\cite{Nilles82,Dine85} for 
supersymmetry breaking. However, before we can make any progress in
phenomenology, two important questions must be answered: is 
supersymmetry broken, and is the dilaton stabilized? Past analyses have 
generally found that, in the absence of a second source of supersymmetry
breaking, the dilaton is destabilized in the direction of vanishing
gauge coupling (the so-called runaway dilaton problem) 
and supersymmetry is unbroken. To address the above questions in
generic linear multiplet models of gaugino condensation, we first 
show how the three issues of supersymmetry breaking, 
gaugino condensation and dilaton stabilization are reformulated, and how they 
are interrelated, by examining the explicit expressions for the gravitino mass 
and the gaugino condensate. A detailed investigation of the vacuum will be 
presented in the following section.

The explicit expression for the gaugino condensate in terms of the dilaton
$\dilaton$ is determined by (3.33):
\be
\bar{u}u\,=\,\frac{1}{e^{2}}\dilaton\mu^{6}
e^{g\,-\,({f+1})/{b\dilaton}}.
\ee
With $g(\dilaton)$=0 and $f(\dilaton)$=0, we recover the result of the simple 
model (3.2) \cite{Binetruy95}. For generic models, the dilaton dependence of 
the gaugino condensate involves $g(\dilaton)$ and $f(\dilaton)$ which 
represent quantum corrections to the tree-level K\"{a}hler potential.
According to our assumption of boundedness for 
$g(\dilaton)$ and $f(\dilaton)$ (especially at $\dilaton$ =0 where 
following (3.12) we have the  boundary 
conditions $g(\dilaton=0)$=0 and $f(\dilaton=0)$=0),
$\dilaton$=0 is the only pole of 
$\:g\,-\,({f+1})/{b\dilaton}.\:$
Therefore, we can draw a simple and clear relation between 
$\langle\bar{u}u\rangle$ and $\langle\dilaton\rangle$: gauginos condense 
(i.e., $\langle\bar{u}u\rangle\neq 0$) if and only if the dilaton is 
stabilized (i.e., $\langle\dilaton\rangle\neq 0$.) 

Another physical quantity of interest is the gravitino mass
$m_{_{\tilde{G}}}$ which is the natural order parameter 
measuring supersymmetry breaking. The expression for 
$m_{_{\tilde{G}}}$ follows directly from $\Lag_{\tilde{G}}$.
\be
m_{_{\tilde{G}}}\,=\,\frac{\,1\,}{\,4\,}b\,
\sqrt{\langle\bar{u}u\rangle},
\ee
where we have used (3.33). This expression for the gravitino mass is 
simple and elegant even for generic linear multiplet models.
From the viewpoint of superstring effective theories, an 
interesting feature of (3.38) is that the gravitino mass 
$m_{_{\tilde{G}}}$ contains no dependence on the modulus $T^{I}$, 
which provides a direct relation between $m_{_{\tilde{G}}}$ and 
$\langle\bar{u}u\rangle$. This feature can be traced to the 
fact that the Green-Schwarz counterterm cancels the $T^{I}$ 
dependence of the superpotential completely, a unique feature of 
the linear multiplet formulation. We recall that, in the chiral 
formulations of gaugino condensation studied previously (with or without the 
Green-Schwarz cancellation mechanism), 
$m_{_{\tilde{G}}}$ always involves a moduli-dependence, 
and therefore the relation between supersymmetry breaking (i.e., 
$m_{_{\tilde{G}}}\neq 0$) and gaugino condensation 
(i.e., $\langle\bar{u}u\rangle\neq 0$) remains undetermined until the
true vacuum can be found. By contrast, in generic linear multiplet 
models of gaugino condensation, there is a simple and 
direct relation, Eq.(3.38): supersymmetry is broken (i.e.,
$m_{_{\tilde{G}}}\neq 0$) if and only if gaugino condensation occurs
($\langle\bar{u}u\rangle\neq 0$). We wish to emphasize that the 
above features of the linear multiplet model are unique in the sense 
that they are simple only in the linear multiplet model. This is related 
to the fact pointed out in Sect. 1 that, once the constraint (2.9) on 
the condensate field $U$ is imposed, the chiral counterpart of the linear 
multiplet model is in general very complicated, and it is more natural to 
work in the linear formulation. Our conclusion of this section is best 
illustrated by the following diagram:

\vspace{0.4cm}
\fbox{\rule[-0.2cm]{0cm}{1.1cm}{\bf\shortstack{Supersymmetry\\Breaking}}}
$\;\Longleftrightarrow\;$
\fbox{\rule[-0.2cm]{0cm}{1.1cm}{\bf\shortstack{Gaugino\\Condensation}}}
$\;\Longleftrightarrow\;$
\fbox{\rule[-0.2cm]{0cm}{1.1cm}{\bf\shortstack{Stabilized\\Dilaton}}}
\vspace{0.4cm}

The equivalence among the above three issues is obvious. Therefore, 
in the following section, we only need to focus on one of the three 
issues in the investigation of the vacuum, for example, the issue
of dilaton stabilization.
\section{Supersymmetry Breaking, Gaugino Condensation
              \mbox{and the Stabilization of the Dilaton}} 
\hspace{0.8cm}\setcounter{equation}{0}
As argued in Sect. 3.1, nonperturbative contributions to the 
K\"{a}hler potential should be introduced to cure the unboundedness 
problem of the simple model (3.2). In the context of the generic model 
(3.10), it is therefore interesting to address the question as to how the 
simple model should be modified in order to obtain a viable theory 
(i.e., with $V_{pot}$ bounded from below). We start with the 
scalar potential $V_{pot}$ arising from (3.30) after solving for the 
auxiliary fields (using (3.33), (3.34) and (3.37)). Recalling that (3.11) 
yields the identity $\:\dilaton\dg+1=1+f-\dilaton\df\:$, we obtain
\be
V_{pot}\,=\,\frac{1}{16e^{2}\dilaton}\{\,
(1+f-\dilaton\df)(1+b\dilaton)^{2}\,-\,3b^{2}\dilaton^{2}\,\}
\mu^{6}e^{g\,-\,({f+1})/{b\dilaton}},
\ee
which depends only on the dilaton $\dilaton$. The necessary and sufficient 
condition for $V_{pot}$ to be bounded from below is
\bea
f-\dilaton\df\,&\geq&\,-\mbox{O}(\dilaton e^{{1}/{b\dilaton}})
\;\;\;\;\;\mbox{for}\;\;\;
\dilaton\,\rightarrow\, 0, \\
f-\dilaton\df\,&\geq&\,\hspace{0.34cm}2\;\;\;\;\;
\hspace{1.1cm}\mbox{for}\;\;\;
\dilaton\,\rightarrow\,\infty.
\eea
It is clear that condition (4.2) is not at all restrictive, and
therefore has no nontrivial implication. On the contrary, condition (4.3) 
is quite restrictive; in particular the simple model violates this condition. 
Condition (4.3) not only restricts the possible forms of the function
$f$ in the strong-coupling regime but also has important
implications for dilaton stabilization and for 
supersymmetry breaking. To make the above statement more 
precise, let us revisit the unbounded potential of Fig.1, with the tree-level 
K\"{a}hler potential defined by $g(V)=f(V)=0$. Adding physically 
reasonable corrections $g(V)$ and $f(V)$ (constrained by 
(4.2-3)) to this simple model should not qualitatively 
alter its behavior in the weak-coupling regime. Therefore, as in
Fig.1, the potential of the modified model in 
the weak-coupling regime starts with $V_{pot}=0$ at $\dilaton=0$,
first rises and then falls as $\dilaton$ increases. On the other 
hand, adding $g(V)$ and $f(V)$ completely alters the 
strong-coupling behavior of the original simple model. As 
guaranteed by condition (4.3), the potential of the modified 
model in the strong-coupling regime is always bounded from below, 
and in most cases rises as $\dilaton$ increases. Joining the 
weak-coupling behavior of the modified model to its 
strong-coupling behavior therefore strongly suggests that
its potential has a non-trivial minimum (at $\dilaton\neq 0$). 
Furthermore, if this non-trivial minimum is global, then the
dilaton is stabilized. We conclude that not only 
does (4.2-3) tell us how to modify the theory, but a 
large class of theories so modified have naturally a stabilized dilaton 
(and therefore broken supersymmetry by the argument of Sect. 3.3). 
In view of the fact that there is currently little knowledge of 
the exact K\"{a}hler potential, the above conclusion, which 
applies to generic K\"{a}hler potentials subject to (4.2--3), is especially 
important to the search for supersymmetry breaking and dilaton
stabilization. Though we are unable to study 
the exact K\"{a}hler potential at present, it is nevertheless
interesting to study models with reasonable K\"{a}hler 
potentials for the purpose of illustrating the significance 
of condition (4.2-3) as well as displaying explicit examples 
with supersymmetry breaking. This will be done in the 
following example.

We start with the consideration of possible 
nonperturbative contributions to the K\"{a}hler potential.
Aside from the Planck scale $M_{P}$, the only natural mass scale in the
theory is the condensation scale $\Lambda_{c}$, that is, the scale at which the 
hidden-sector gauge interaction becomes strong. As is 
well known, it follows from the renormalization group equation for the running 
of the gauge coupling that $\Lambda_{c}$ depends exponentially on the dilaton 
$\dilaton$ as $\;\Lambda_{c}\,\sim\,e^{-\,{1}/{6b\dilaton}}$, 
which is consistent with the results of the simple 
model in Sect. 3.1. Therefore, on dimensional 
grounds, the field-theoretical nonperturbative contribution 
to the K\"{a}hler potential has the generic form 
$\,\,V^{-m}e^{-\,{n}/{6bV}}/M_{P}^{n-2}\,\,$ ($M_{P}$=1 
in our convention), where $n\geq 2$ and $m\geq 0$ 
\cite{Derendinger95}. In the following example, we consider the
leading-order nonperturbative contribution ($n=2$ and $m=0$) 
to the K\"{a}hler potential:
\be
f(V)\,=\,A_{_{f}}e^{-\,{1}/{3bV}},
\ee
where $A_{_{f}}$ is a constant to be determined by the 
nonperturbative dynamics. The regulation conditions (4.2-3)
require $A_{_{f}}\geq 2$. In Fig. 2, $V_{pot}$ is plotted versus
the dilaton $\dilaton$, where $A_{_{f}}=6.92$ and $\mu$=1. Fig. 2 has two
important features. First, $V_{pot}$ of this modified theory is 
indeed bounded from below, and the dilaton is stabilized. Therefore, 
we obtain supersymmetry breaking, gaugino condensation and 
dilaton stabilization in this example. The gravitino mass is
$m_{_{\tilde{G}}}\,=\,7.6\times 10^{-5}$ in Planck units. 
Secondly, the $vev$ of dilaton is stabilized at the phenomenologically 
interesting range ($\langle\dilaton\rangle\,=\,0.45\,$ in Fig. 2).
Furthermore, the above features involve no unnaturalness since they 
are insensitive to $A_{_{f}}$. Fig. 2 is a nice realization 
of the argument in the preceding paragraph. It should be contrasted
with the racetrack models where at least three gaugino condensates
and large numerical coefficients are needed in order to achieve 
similar results. We can also consider possible stringy 
nonperturbative contributions to the K\"{a}hler potential 
suggested in \cite{Shenker90}. It turns out that we obtain the 
same general features as those of Fig. 2. This is not surprising
since, as argued in the preceding paragraph, the important features 
that we find in Fig. 2 are common to a large class of models. 

Note that the value of the cosmological constant is irrelevant to the 
arguments presented here and in Sect. 3.3. In other words, 
the generic model (3.10) suffers from the usual cosmological constant problem,
although we can find a fine-tuned subset of models whose cosmological 
constants vanish. For example, the cosmological constant of Fig. 2 
vanishes by fine tuning $A_{_{f}}$. It remains an open question as to 
whether or not the cosmological constant problem could be resolved within 
the context of the linear multiplet formulation of gaugino condensation if 
the exact K\"{a}hler potential were known.
\section{Concluding Remarks}
\hspace{0.8cm}\setcounter{equation}{0}

    We have presented a concrete example of a solution to the infamous runaway 
dilaton problem, within the context of local supersymmetry and the linear 
multiplet formulation for the dilaton.  We considered models for a static 
condensate that reflect the modular anomaly of the effective field theory while
respecting the exact modular invariance of the underlying string theory.  The 
simplest such model~\cite{Binetruy95,sduality} has a nontrivial potential that 
is, however, unbounded in the direction of strong coupling. Including 
nonperturbative corrections~\cite{Derendinger95,Shenker90} to the K\"ahler
potential for the dilaton, the potential is stabilized, allowing a vacuum
configuration in which condensation occurs and supersymmetry is
broken.  This is in contrast to previous analyses, based on the
chiral formulation for the dilaton, in which supersymmetry breaking with a
bounded vacuum energy was achieved only by introducing an additional source of
supersymmetry breaking, such as a constant term in the superpotential
\cite{chiral91,Dine85,Nilles95}.

In further contrast to most chiral models studied, supersymmetry breaking
arises from a nonvanishing vacuum expectation value of the auxiliary field
associated with the dilaton rather than the moduli: roughly speaking, in the
dual chiral formulation, $\langle F_S \rangle\ne 0$ rather than
$\langle F_{T}^{I} \rangle\ne 0$.  As a consequence, gaugino masses and 
A-terms are generated at tree level. Although scalar masses are still 
protected at tree level by a Heisenberg symmetry~\cite{heis}, 
they will be generated at one loop by renormalizable interactions. 
For the model considered here, the hierarchy (about five orders of magnitude) 
between the Planck scale and the gravitino mass is insufficient to account for 
the observed scale of electroweak symmetry breaking. A possible 
avenue for improving this result is to consider multiple gaugino
condensation;  in realistic orbifold compactifications the hidden gauge group
${\cal G}$ is in general a product group: ${\cal G} = \Pi_a{\cal G}_a$. The 
generalization of our formalism to the multi-condensate case will be 
considered elsewhere. 

The Kalb-Ramond field (or the axion, in the dual description) remains
massless in the static models considered here, and therefore we still need
to explain how the axion mass can be generated. It has recently been shown 
in the context of global supersymmetry \cite{Binetruy95} that a mass term 
for the axion is naturally generated if kinetic terms for $U$ and $\bar{U}$
are included. It is therefore worth studying the extension of this paper
to the nonstatic case.  Consider the following generic
linear multiplet model with a single dynamical condensate:
\bea
K\,&=&\,\ln V\,+\,g(V,\bar{U}U)\,+\,G, \nonumber \\
\Lag_{eff}\,&=&\,\superint\,E\,\{\,(\,-2\,+\,f(V,\bar{U}U)\,)\,+\, 
bVG\,+\,bV\ln(e^{-K}\bar{U}U/\mu^{6})\,\}.\hspace{1.5cm}
\eea
The model defined by (5.1) is a straightforward generalization of (3.10), 
where the 
quantum corrections to the K\"{a}hler potential, $g$ and $f$, are now taken
to be functions of $\bar{U}U$ as well as of $V$. The construction of the 
component lagrangian for the nonstatic model (5.1) is similar to that for the
static model (3.10) presented in Sect. 3.2. For example, the condition for a 
canonical Einstein term for the generic nonstatic model turns out to be:
\be
(\,1\,+\,Z\frac{\partial f}{\partial Z}\,)
(\,1\,+\,V\frac{\partial g}{\partial V}\,)\,=\,
(\,1\,-\,Z\frac{\partial g}{\partial Z}\,)
(\,1\,-\,V\frac{\partial f}{\partial V}\,+\,f\,),
\ee
where $Z\equiv\bar{U}U$. It is clear that (3.11) is the static limit of (5.2),
where $g$ and $f$ are independent of $\bar{U}U$. As suggested by terms that
arise both from string corrections~\cite{Ant} at the classical level and 
from field-theoretical loop corrections~\cite{kamran}, we have studied 
the nonstatic model with generic functions $g$ and $f$ that are s-duality 
invariant in the sense defined in~\cite{sduality}. That is, $g$ and $f$ are 
functions only of the  s-duality invariant superfield variable 
$\bar{U}U/V^{2}$. It turns out that the scalar potential $V_{pot}$ of the 
nonstatic model with s-duality invariance is always unbounded from below in the
strong-coupling limit $\dilaton\,\rightarrow\,\infty$. The origin of this
unboundedness problem is similar to that of the simple static model studied in 
Sect. 3.1, and again it reflects the absence of nonperturbative 
contributions to the K\"{a}hler potential. We expect that the
unboundedness problem of the nonstatic model will be cured when 
nonperturbative contributions to the K\"{a}hler potential are included. 
Studies along this line are in progress.
\section*{Acknowledgements}
\hspace{0.8cm}
PB and MKG would like to acknowledge the hospitality of the Santa Barbara 
Institute for Theoretical Physics, where this work was initiated.
This work was supported in part by the Director, Office of 
Energy Research, Office of High Energy and Nuclear Physics, Division of 
High Energy Physics of the U.S. Department of Energy under Contract 
DE-AC03-76SF00098 and in part by the National Science Foundation under 
grant PHY-95-14797.
\pagebreak

\pagebreak
\clearpage
\hspace{1.6in} FIGURE CAPTIONS

\vskip 0.5in
Fig.1: The scalar potential $V_{pot}$ (in Planck units)
is plotted versus the dilaton $\dilaton$. $\mu$=1.

\vskip 0.5in
Fig.2: The scalar potential $V_{pot}$ (in Planck units)
is plotted versus the dilaton $\dilaton$. $\,A_{_{f}}=6.92\,$ and
$\,\mu$=1.


\begin{thebibliography}{99}

\bibitem{Linear} \mbox{S. Ferrara and M. Villasante,
\mbox{Phys. Lett. {\bf B186}} (1987) 85;} \linebreak
\mbox{P. Bin\'{e}truy, G. Girardi, R. Grimm and M. M\"{u}ller, 
Phys. Lett.} \mbox{{\bf B195} (1987) 389;}
\linebreak \mbox{S. Cecotti, S. Ferrara and M. Villasante,
Int. J. Mod. Phys. {\bf A2} (1987)} 1839; \linebreak
\mbox{S. Ferrara, J. Wess and B. Zumino,
Phys. Lett. {\bf B51} (1974) 239}.
\bibitem{Gaillard92} M.K. Gaillard and T.R. Taylor, 
\mbox{Nucl. Phys. {\bf B381}} (1992) 577.
\bibitem{KL} V.S. Kaplunovsky and J. Louis, Nucl. Phys. {\bf B444} (1995) 191.
\bibitem{frad} E.S. Fradkin and A.A. Tseytlin, Ann. Phys. {\bf 162}
(1985) 31.
\bibitem{Burgess95}\mbox{C.P. Burgess, J.-P. Derendinger, 
F. Quevedo and M. Quir\'{o}s,}
\mbox{Phys. Lett. {\bf B348}} (1995) 428.
\bibitem{Binetruy95} P. Bin\'{e}truy, M.K. Gaillard and 
T.R. Taylor, \mbox{Nucl. Phys. {\bf B455}} (1995) 97.
\bibitem{sduality}P. Bin\'etruy and M.K. Gaillard, Phys.\ Lett.
{\bf B365} (1996) 87.
\bibitem{Binetruy90} P. Bin\'{e}truy, G. Girardi, R. Grimm
and M. M\"{u}ller, \mbox{Phys. Lett. {\bf B189}} \mbox{(1987) 83;}
\linebreak \mbox{P. Bin\'{e}truy, G. Girardi and R. Grimm,} 
preprint \mbox{LAPP-TH-275/90}.
\bibitem{Binetruy91} P. Bin\'{e}truy, G. Girardi, R. Grimm 
and M. M\"{u}ller, \mbox{Phys. Lett. {\bf B265}} (1991) 111.
\bibitem{Dixon90} L.J. Dixon, V.S. Kaplunovsky and 
J. Louis, Nucl. Phys. {\bf B355} (1991) 649.
\bibitem{Derendinger92} J.-P. Derendinger, S. Ferrara,
C. Kounnas and F. Zwirner, \mbox{Nucl. Phys. {\bf B372}} (1992) 145.
\bibitem{Ovrut93} G.L. Cardoso and B.A. Ovrut, 
\mbox{Nucl. Phys. {\bf B392}} (1993) 315.
\bibitem{Giveon89} A. Giveon, N. Malkin and 
E. Rabinovici, Phys. Lett. {\bf B220} (1989) 551.
\bibitem{ant} I. Antoniadis, K.S. Narain and T.R. Taylor, 
Phys. Lett. B267 (1991) 37.
\bibitem{Adamietz93} P. Adamietz, P. Bin\'{e}truy, G. Girardi
and R. Grimm, \mbox{Nucl. Phys. {\bf B401}} (1993) 257.
\bibitem{Derendinger94}\mbox{J.-P. Derendinger, F. Quevedo
and M. Quir\'{o}s, Nucl. Phys. {\bf B428}} (1994) 282.
\bibitem{Pillon}\mbox{P. Bin\'{e}truy, F. Pillon,
G. Girardi and R. Grimm,} preprint hep-th/9603181, 
LPTHE-Orsay 95/64, ENSLAPP-A-553/95, CPT-95/P.3258.
\bibitem{3form} \mbox{G. Girardi and R. Grimm,
Phys. Lett. {\bf B260} (1991) 365;} \linebreak
\mbox{S.J. Gates, Nucl. Phys. {\bf B184} (1981) 381}.
\bibitem{Veneziano82} G. Veneziano and S. Yankielowicz,
\mbox{Phys. Lett. {\bf B113}} (1982) 231; \linebreak
     \mbox{T.R. Taylor,
\mbox{Phys. Lett. {\bf B164}}, (1985) 43;} \linebreak
     \mbox{P. Bin\'{e}truy and M.K. Gaillard, 
\mbox{Phys. Lett. {\bf B232}} (1989) 82;} \linebreak
\mbox{S. Ferrara, N. Magnoli, T.R. Taylor and G. Veneziano,
Phys. Lett.} \mbox{{\bf B245} (1990) 409}.
\bibitem{chiral91} P. Bin\'{e}truy and M.K. Gaillard, 
\mbox{Nucl. Phys. {\bf B358}} (1991) 121.
\bibitem{Derendinger95}\mbox{T. Banks and M. Dine,
Phys. Rev. {\bf D 50} (1994) 7454;} \linebreak
C.P. Burgess, J.-P. Derendinger, 
F. Quevedo and \mbox{M. Quir\'{o}s}, preprint 
\mbox{hep-th/9505171}, \mbox{CERN-TH/95-111}.
\bibitem{Shenker90} S.H. Shenker, in {\it Random Surfaces
and Quantum Gravity}, Proceedings of the NATO Advanced Study
Institute, Cargese, France, 1990, edited by O. Alvarez, 
E. Marinari, and P. Windey, NATO ASI Series B: Physics Vol.262
(Plenum, New York, 1990).
\bibitem{Wess} J. Wess and J. Bagger, Supersymmetry and 
Supergravity, Princeton Series in Physics (Princeton U.P.,
Princeton, 1992).
\bibitem{Nilles82}\mbox{H.P. Nilles, 
Phys. Lett. {\bf B115} (1982) 193;} \linebreak
S. Ferrara, L. Girardello and H.P. Nilles,
Phys. Lett. {\bf B125} (1983) 457. 
\bibitem{Dine85} M. Dine, R. Rohm, N. Seiberg and E. Witten,
Phys. Lett. {\bf B156} (1985) 55.
\bibitem{heis} P.\ Bin\'etruy and M.K. Gaillard, 
Phys.\ Lett.\ {\bf B195} (1987) 382.
\bibitem{Nilles95} G.D. Coughlan, G. Germain, G.G. Ross and 
G. Segr\'{e}, Phys. Lett. {\bf B198} (1987) 467;\newline
P. Bin\'etruy and M.K. Gaillard, Phys. Lett. {\bf B253} (1991) 119;
\newline Z. Lalak, A. Niemeyer and H.P. Nilles,
\mbox{Nucl. Phys. {\bf B453}} (1995) 100.
\bibitem{Ant} I. Antoniadis, E. Gava, K.S. Narain and T.R. Taylor,
Nucl. Phys. {\bf B432} (1994) 187.
\bibitem{kamran} M.K.~Gaillard, V.~Jain and K. Saririan, LBL-34948, 
preprint in preparation.
\end{thebibliography}
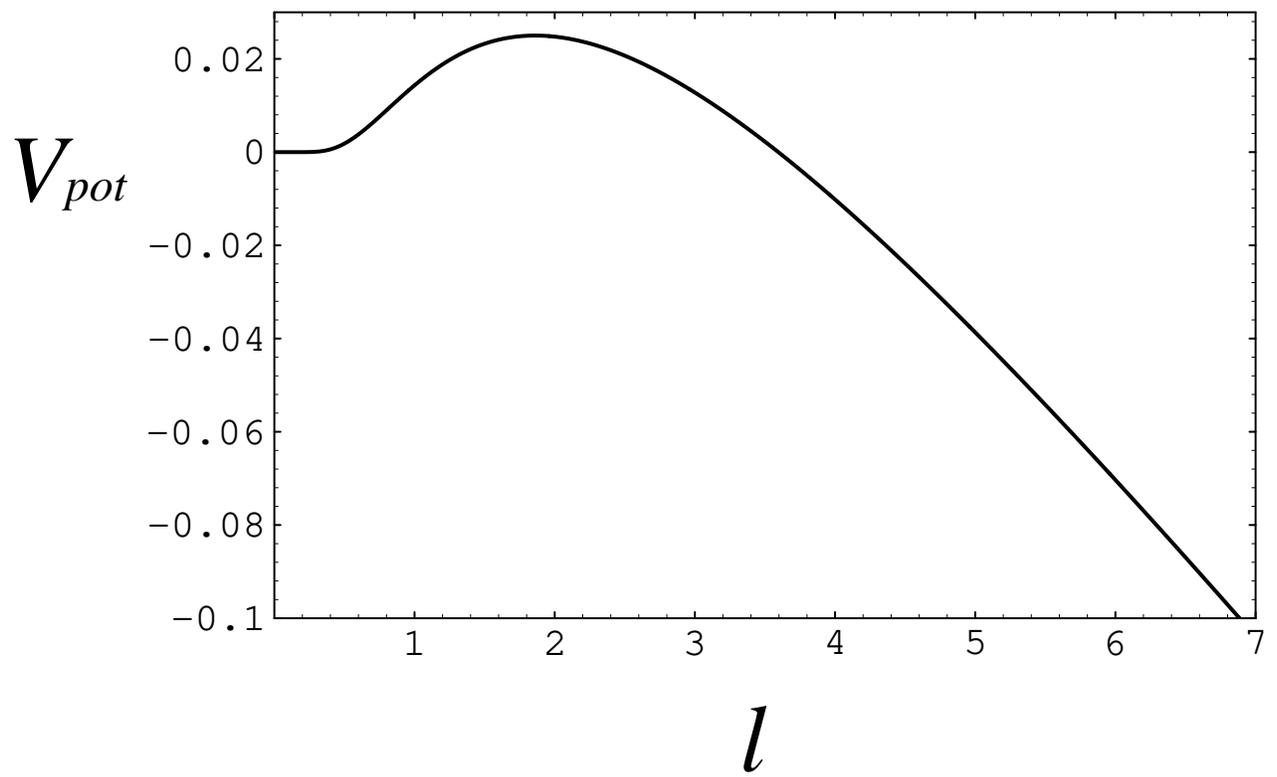
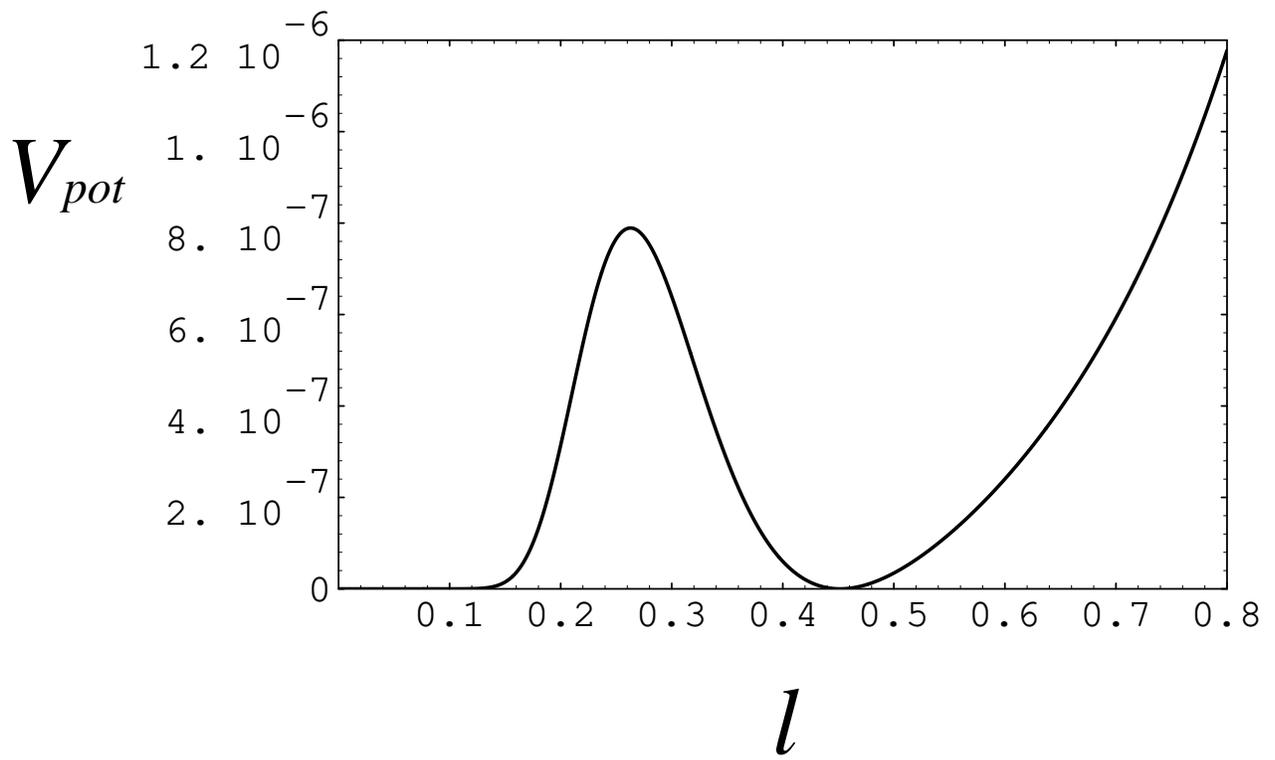
\end{document}